\documentclass[prodmode,acmtecs]{acmsmall} 

\usepackage{subcaption}
\usepackage{soul}

\usepackage{listings}
\usepackage{array}
\usepackage{rotating}
\usepackage{textcomp}
\usepackage{enumitem}
\usepackage{multirow}
\usepackage{floatrow}
\usepackage{float}
\usepackage{slashbox,pict2e}
\newfloatcommand{capbtabbox}{table}[][\FBwidth]
\usepackage{graphicx, epstopdf}
\DeclareGraphicsExtensions{.ps}
\usepackage{caption}
\usepackage[ruled]{algorithm2e}
\renewcommand{\algorithmcfname}{ALGORITHM}
\SetAlFnt{\small}
\SetAlCapFnt{\small}
\SetAlCapNameFnt{\small}
\SetAlCapHSkip{0pt}
\IncMargin{-\parindent}

\begin{document}


\title{SEAT: A Taxonomy to Characterize Automation in Software Engineering}
\author{Shipra Sharma
\affil{Indian Institute of Technology Ropar, Rupnagar, India}
Balwinder Sodhi
\affil{Indian Institute of Technology Ropar, Rupnagar, India}}

\begin{abstract}
Reducing cost and time required to build high quality software is a major goal for software developers. Building tools and techniques that can help achieve such a goal is the chief aim for Automated Software Engineering (ASE) researchers. However, in order to be effective an ASE researcher or professional must understand the \emph{characteristics} of both \emph{successful} and not-so-successful ASE tools, and the constituent techniques employed by such ASE tools.

In this paper we present such a characterization of ASE tools and major constituent techniques from different areas of computer science and engineering that have been employed by such ASE tools. To develop the characterization we carried out an extensive systematic literature review over about 1175 ASE research articles. One of our key goal was to identify useful relationships/patterns among ASE tools, their constituent techniques and the software development life cycle (SDLC) activities that these tools targeted.

For example, we observed that the predominantly used constituent techniques can be classified into 11 categories. Only $\approx$26\% ASE tools (\emph{from our sample}) leveraged more than one constituent techniques to achieve their goal. We also observed that a significant number ($\approx$63\%) of ASE tools did not have much \emph{impact}. More than 50\% of the sampled ASE tools targeted Testing and Verification activities possibly implying the ease of automation there. In terms  of changes in popularity of constituent techniques with time we did not observe any clear trend.

We organized the results of our characterization as a taxonomy called SEAT (Software Engineering Automation Taxonomy). A salient feature of SEAT is that it focuses on automation of activities from all phases of SDLC. Such a taxonomy, among other applications, shall enable synthesizing new automation tools for different SDLC activities. Re-composing existing systems to achieve better features will also be possible. Further, the taxonomy has been realized as a graph database using neo4j\textregistered ~ (an open source graph database), which can be queried using an SQL like language.
The graph database allowed us to uncover hidden relationships by way of exhaustive search for connections and paths between different nodes (i.e. concepts). We demonstrate the efficacy of SEAT by discussing few practical use cases. 

\end{abstract}
%
%






\maketitle

\section{Introduction}
With ever growing adoption of IT and software systems, building quality software in shorter time and at lower cost has been a major goal for software engineering professionals. 
Automated Software Engineering (ASE) community over the years has made significant contribution towards achieving this goal by developing tools and techniques to automate various activities of different phases of Software Development Life Cycle (SDLC). 
Such automation has helped in reducing the cost and time required to build good quality software. Almost every fundamental activity in SDLC has seen some type of automation. These activities typically are requirement engineering, design, implementation, testing and product's maintenance \cite{Sommerville2010}. 

Though, life cycle of software development can follow varied methodologies, the basic activities in each methodology remain more or less the same. For instance, whether one employs the Waterfall, V or Spiral model, the phases such as requirements analysis, architecture, design, testing and so on are always involved at some level in software development. 
Each phase has its own set of automation tools and techniques (henceforth called \textbf{ASE tools}) which are intended for reducing the manual work in respective phases. In order to automate tasks in the targeted SDLC activities such ASE tools leverage and depend on techniques (henceforth called \textbf{constituent techniques}) from several other areas of computing such as formal methods, semantic computing, natural language processing, information retrieval, knowledge representation and so on. 


For example, many earlier ASE tools employed formal methods while several of the recent
ones rely on semantic computing techniques. Diversity in constituent techniques on which such ASE tools are built leads to varying levels of quality, adoption and success for those ASE tools. 

From the perspective of end users, researchers and academicians who are engaged in using, developing and teaching about ASE tools it is important to understand the whole landscape of ASE tools and techniques, underlying design approach of each tool and many other related factors.


%

However, even after many years and a great deal of advancements in automation of tasks in SDLC phases, it is difficult to find in literature a methodical and SDLC-centric characterization of these automation tools and techniques. Though there are few studies, such as \cite{rafi2012benefits}, \cite{anand2013orchestrated} that examin testing phase, \cite{gulwani2010dimensions} that examins works in program synthesis etc., they all address only a subset of SDLC activities and none of them considers the entire SDLC.

We carried out a systematic state-of-the-art literature review of the automation tools and techniques developed by ASE researchers for various phases of SDLC. A major aim of our study was to identify useful relationships and patterns if any existed among ASE tools, their constituent techniques and the SDLC activities that these tools targeted.

Based on our findings from this study, we synthesized a non-orthogonal taxonomy -- that we call Software Engineering Automation Taxonomy (SEAT) -- that characterizes ASE tools, their constituents techniques and the SDLC activities that they target. We implemented this taxonomy in a graph database for easy querying along different dimensions. A graph database allowed us to uncover hidden relationships by way of exhaustive search for connections and paths between different nodes (i.e. concepts).


Rest of this paper is structured as follows. In Section-\ref{sec:res_method} we describe our overall research methodology; we outline our key research questions in this section. Section-\ref{sec:data_ext} presents our approach for eliciting relevant information from research literature. A detailed discussion, along our research questions, of the patterns we observed in ASE research literature is presented in Section-\ref{sec:results_discussion}. We then leverage these observations to develop a taxonomy for automation in software engineering in Section-\ref{sec:seat}. Finally we conclude the paper in Section-\ref{sec:conclusions}. 

%
\section{Research Method}
\label{sec:res_method}
In order to synthesize our taxonomy we first carried out a Systematic Literature Review (SLR) of the ASE domain.
We used a tool called \emph{StArt} (\cite{fabbri2012managing}, \cite{fabbri2016improvements}) to study and analyze the ASE research articles for various phases of SDLC. This review and analysis of data was done as per the guidelines outlined in \cite{keele2007guidelines}. We selected StArt after examining various SLR tools \cite{alreview}. After trying few of these tools we found StArt to be the most user-friendly and following all the steps prescribed for conducting an SLR. StArt assists in a step wise paper selection and extraction process as mentioned in \cite{keele2007guidelines}. 

Based on certain pre-defined keywords and parameters (discussed shortly), the StArt tool allowed us to catalogue all the ASE contributions (i.e. ASE tools and techniques) at one place, and also helped in analysis and visualization of the information along different dimensions.

\subsection{Research Questions}
\label{sec:res_questions_list}
A contextual understanding of the relationship of various aspects of ASE tools with the constituent techniques is essential for identifying and understanding the limitations and opportunities in the domain of ASE. To address the existing gaps in such an understanding is one of the motivation factors for us to review state of the art for ASE tools meant for all SDLC phases. Our main objective was to determine whether there exists any interesting patterns in respect of: 
\begin{enumerate}
\item relationships between the ASE tools and the issues/activities that these tools addressed.
\item relationships between the constituent techniques and the ASE tools that leveraged them.
\end{enumerate}
Another key issue that we intend to address is related to methodical selection of: 
\begin{enumerate}
\item an ASE tool by a practitioner who is looking to find a best fit for his/her needs, and
\item constituent techniques by ASE researcher who is looking to develop the next ASE tool.
\end{enumerate}

\begin{figure}
\CenterFloatBoxes
\begin{floatrow}
\ffigbox{%
  {\includegraphics[scale=0.28]{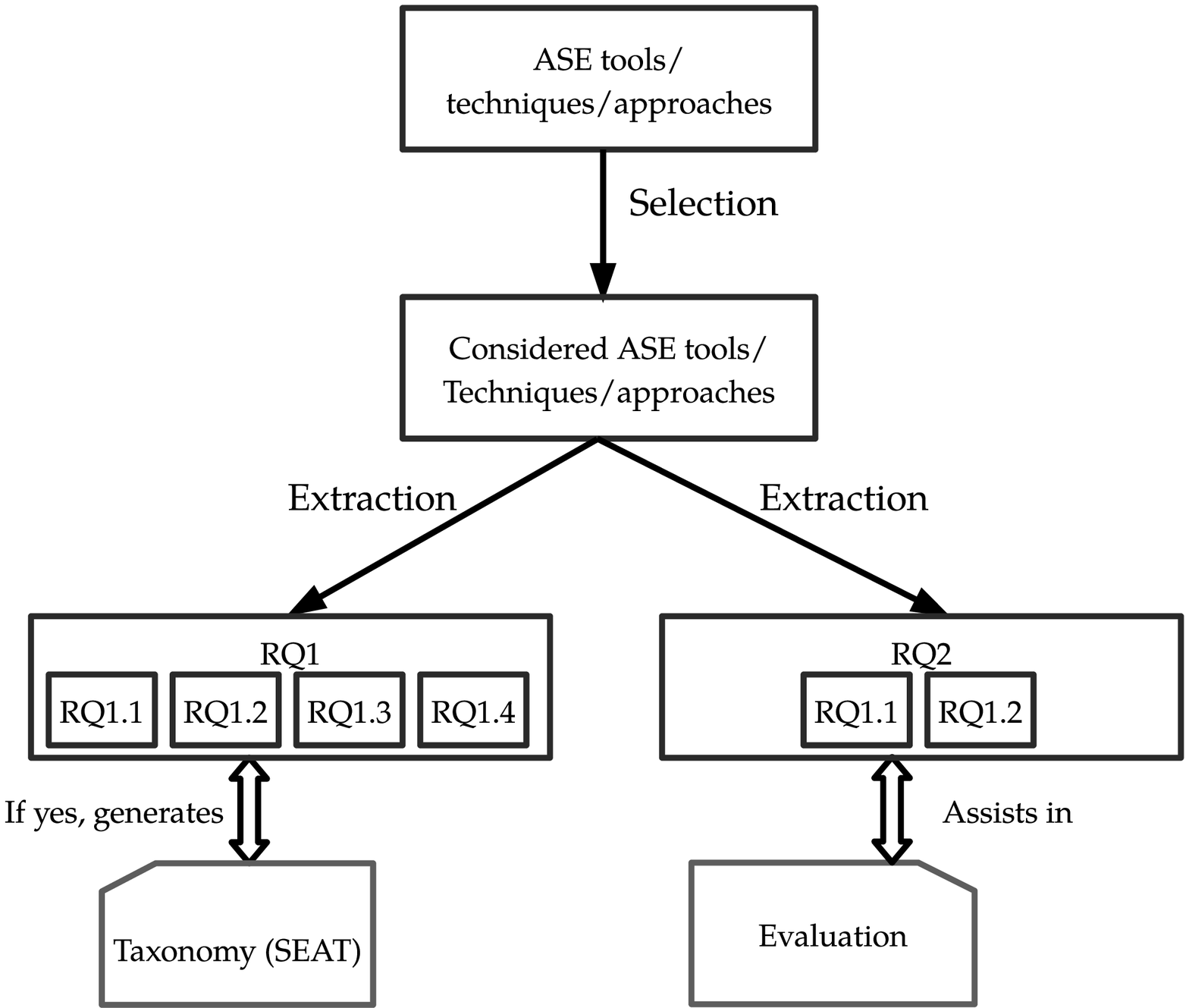}}
}{%
  \caption{Research Model}%
  \label{mymodel}
}
\killfloatstyle
\ttabbox{
  \renewcommand{\arraystretch}{1.2}
  \begin{tabular}{p{2.2cm} p{4cm}}  \hline \hline
  Population &  Software Engineer (may be in one or more role of architect, designer, developer, tester)\\ \hline 
  Intervention & Tools and Technologies to perform (semi) automation in fundamental software process development activity.\\ \hline
  Comparison & None\\ \hline
  Outcome & SEAT (Taxonomy)\\ \hline
  Context & Software Engineering and Automation \\ \hline \hline
  \end{tabular}
}{
 \caption{PICOC Criteria}
 \label{PICOC-TAB}
}
\end{floatrow}
\end{figure}

The above motivating factors led us to more descriptive research questions which we describe shortly. These research questions have been framed according to the five criteria recommended by Kitchenham et al. \cite{keele2007guidelines} i.e., Population, Intervention, Comparison, Outcome and Context (PICOC). The PICOC criteria for our SLR is depicted in Table-\ref{PICOC-TAB}.

The key research questions that we explored are as follows:
\begin{enumerate}[label={RQ.}{\arabic*}: ]
	\item Are there any patterns of relationships between SDLC activities, the ASE tools and the constituent techniques.
	\begin{enumerate}[label={\arabic*}. ]
		\item What are the various constituent techniques that are used to automate different phases of SDLC? In other words, what methods and techniques from different areas of computing are leveraged when constructing automation tools of SDLC?
		\item Are these constituent techniques ``SDLC activity specific"? In other words, does a particular constituent techniques find more acceptance in a particular activity of SDLC?
		\item Does a particular constituent technique span across more than one activity of SDLC?
		\item Have these constituent techniques gained/lost acceptance with time?
	\end{enumerate}
	\item To what extent does the automation of a particular phase takes place? Does this extent depend on that phase?    
	\begin{enumerate}[label={\arabic*}. ]
		\item Is there a particular SDLC activity which is more difficult/easy to automate?
		\item Is level of automation constrained by availability of constituent techniques?
	\end{enumerate}

\end{enumerate}

Fig. \ref{mymodel} depicts our SLR model and its expected outputs. 


\subsection{Search Strategy}
In order to find answers for our research questions we fetched relevant research literature. For this we first identified suitable sources/repositories of such research literature. Next, we formulated search strings in order to filter the most relevant articles from those sources. Details of these steps are as follows.

\subsubsection{Repositories Searched}
Due the vast amount of literature that exists in the ASE domain we restricted ourselves to ACM portal. This choice was due to following reasons: All major conferences in automation or software (e.g. Automated Software Engineering, International Conference on Software Engineering) are indexed by ACM. All journals of ACM related to software are highly ranked. 

Also, if an ASE tool got improved in another paper (not indexed by ACM), we added the paper which discussed the improved version of that ASE tool. Finally, so as not to miss any important contribution we also added all those relevant paper which were cited in the selected papers (known as reference chaining \cite{achimugu2014systematic}). While searching for literature on specific repositories we used suitable search options which allowed multiple keyword searches. No restriction on the date of publishing or number of citations or any other constraint was put. Everything that was found via such search queries was considered for examining and analysis.

\subsubsection{Composition of search queries}
\label{searchstr}
We formulated a template search query in which keyword parameters were dynamically set based on the SDLC phase for which we wanted to retrieve the ASE tools literature. Search queries were also executed for: a) synonyms of the given keyword/concept, and b) multiple keywords connected by booleans operators such as \texttt{AND, OR}.

\subsection{Paper Selection}
\label{PaperselSect}
Our primary search elicited $1175$ papers. These papers were then filtered based on title and abstract analysis. For each of these papers we went through their abstract and title and removed those which did not match the inclusion criteria (Table-\ref{Papernum}). The inclusion criteria is depicted in Table-\ref{IE}. As a result we were left with $326$ relevant papers. The number of papers was greatly reduced due to (among other reasons) duplicated and irrelevant matches that search queries returned. For example, keyword `automation' had also matched papers related to `automation software industry'. 

\begin{table}
\RawFloats
\centering
\setlength\tabcolsep{4pt}
\begin{minipage}{0.45\textwidth}
\centering
\tbl{Paper Identification And Filtering Process}{
\centering
\begin{tabular}{p{5cm} r}
\hline\hline
\textbf{Process description} & \textbf{\# of papers} \\ \hline
Papers obtained in primary search & 1175 \\
After Filtering Based on Title and Abstract Analysis & 326 \\
Papers obtained from Secondary Methods & 19 \\
Papers Rejected After Complete Reading & 179 \\
Papers Accepted & 164 \\
\hline
\end{tabular}}
\label{Papernum}
\end{minipage}%
\hfill
\begin{minipage}{0.45\textwidth}
\centering
\tbl{Quality Criteria}{
\centering
\begin{tabular}{p{4cm} r}
\hline\hline
\textbf{Parameter} & \textbf{Applicability area} \\ \hline
Is the aim of the research/article clearly articulated & Generic\\
Are the results mentioned in paper credible & Generic\\
Is the software development phase clearly mentioned & SDLC\\
Does the tool$/$technique in practice decrease/replace any manual effort & Automation \\
\hline
\end{tabular}}
\label{Qual}
\end{minipage}
\end{table}

\begin{table}
\tbl{Inclusion and Exclusion Criteria}{
\centering
\begin{tabular}{p{0.3\textwidth} | p{0.7\textwidth}}
\hline\hline
\textbf{Include a paper if:} & \textbf{Exclude a paper if:} \\ \hline
It described automation of any SDLC activity. & It presents merely an empirical study about an ASE problem, or\\
& It is not related to automation of an SDLC activity, or \\
& It describes automation in hardware platforms, even if via software.\\
\hline
\end{tabular}}
\label{IE}
\end{table}

We then added papers from secondary referencing. That is, relevant literature references that were present in the set of papers resulting from narrowed search. Also, any tool/technique which was proposed in the selected $326$ papers but had an improved version elsewhere was added. This lead to addition of $19$ more papers to the overall set selected. These $326 + 19 = 345$ papers were analyzed by complete reading. It was checked whether they really satisfied the inclusion criteria and related to our research questions. We chose a relatively narrow inclusion criteria because we wanted to select only those papers which fully/semi-automated one or more of the SDLC phases. We rejected papers which: (i) Did not propose a tool/technique/approach for automation, or (ii) If a paper provided an automation technique which did not explicitly assist in any SDLC phase (for example ``automation in structured data management''), or (iii) If the automation is specific to some particular computing platform requirements like ``automation for parallel distributed system". These checks further led to removal of $179$ papers resulting in only $164$ articles. We assigned a unique ID (\texttt{P1}  through \texttt{P164}) to each of these papers for easy referencing. These IDs and other meta-data of all these papers can be accessed at: \url{https://goo.gl/DnJvWy}. In subsequent sections we refer to a paper by its unique ID, e.g. as [P12].

\subsection{Assessment of Relevance of Selected Papers}
In this section we discuss how papers were assessed for their relevance for our study. This quality assessment was inspired by quality check idea provided in \cite{keele2007guidelines}. We assessed each paper in terms of whether it had adequate information to answer our research questions. The quality criteria is depicted in Table-\ref{Qual}. For each parameter we marked the paper with ``Yes" or ``No". A Paper having more than $3$ ``No" was discarded. As all the selected papers had less than $3$ ``No", none was rejected in this step.

\section{Information Extraction from Research Literature}
\label{sec:data_ext}
To address our research questions, we created a data extraction form in StArt tool. Our main objective here was to scrutinize each of the $164$ papers individually and gather information to have an overall analysis of automation in SDLC. To this end we extracted two types of information from all these $164$ papers:

\begin{enumerate}
\item \textbf{Meta-data}: Bibliographic information (such as title, author), a unique id, journal/conference details. Table-\ref{metadatatab} depicts the complete list.
\item \textbf{Qualitative Information}: Information specific to our research questions. This was initially done on the basis of what the articles reported. Table-\ref{Qualdatatab} depicts entire list of qualitative features on which this data is based.  
\end{enumerate} 
The extraction was conducted by the first author. This was entered in the StArt tool and shared with the second author. For validation, the second author checked few randomly selected papers on two parameters: 
\begin{enumerate}
\item Application of inclusion/exclusion criteria to a paper.
\item The information that was extracted from a paper.
\end{enumerate}

Any anomalies or differences in opinion were discussed. There were no differences in the second parameter (due to the very specific nature of questions in the extraction form). Although, in case of first parameter there were few differences due to difference in opinion of whether the work can be considered automation of task/s in SDLC phase. A consensus about such papers was reached by dividing each SDLC phase into a sub-phases (discussed in detail in Section-\ref{tax} while discussing the proposed taxonomy, SEAT). If any sub-phases were being automated then the paper was included, otherwise it was excluded. All those papers that finally met the selection criteria were then included because they were found to be automating one or the other sub-phase of a phase of SDLC.

\begin{table}
\tbl{Extracted Meta-data}{
\centering
\begin{tabular}{p{0.35\textwidth} | p{0.65\textwidth}}
\hline\hline
\textbf{Data item} & \textbf{Description} \\ \hline
PID & Each Paper has unique identifier generated automatically by StArt\\ \hline
Title, Name, Author Details, Year & Extracted from individual paper\\ \hline
Article Type & Whether a conference, journal paper or a dissertation\\ \hline
Keywords & In addition to those present in paper, also considered the ones added by us to help in further classification\\ \hline
Type of work & Whether a working tool, a theoretical approach or extension of existing tool \\
\hline
\end{tabular}}
\label{metadatatab}
\end{table}

\begin{table}
\tbl{Extracted Qualitative Information}{
\centering
\begin{tabular}{p{0.20\textwidth} | p{0.40\textwidth} | p{0.40\textwidth}}
\hline\hline
\textbf{Data item} & \textbf{Description} & \textbf{Why extracted}\\ \hline
SDLC phases automated & The number and name of SDLC phases that were semi-/automated by the paper &  Its impact on our synthesized Taxonomy\\ \hline
Degree of Automation & Does the paper provide complete automation or semi-automation & This is will decide the extent to which manual work has decreased. \\ \hline
How was automation done? &  Which technique/s were used to achieve this automation & Identify if there were any patterns in automating a phase with specific methodologies? \\ \hline
Degree of user-effort & Does it require the user to learn some new tool /technology & Is automation really decreasing manual effort?\\ \hline
Usage & Does any software (open-source or otherwise) use the proposed tool or methodology? & Has the tool found a practical or inspirational use in other such tools etc.\\ \hline
\hline
\end{tabular}}
\label{Qualdatatab}
\end{table}

\section{Discussion of Results in Context of Research Questions}
\label{sec:results_discussion}
In order to infer useful patterns and insights from these papers we extracted from these papers all such information which is related to research questions of our study. 
Further, to develop a systematic understanding of automation in different phases of SDLC, and to synthesize SEAT taxonomy we determined the pair-wise correlation among different types of information which we extracted from the selected papers. For example, the correlation between SDLC phases and the constituent techniques used by the ASE tools targeted for respective SDLC phase. Our findings from such analysis are discussed next. We first present the interesting patterns and related information which was observed. Then we discuss these observations in context of our research questions.

(\emph{For a paper specific detailed discussion around individual findings you may refer to the ``extra$\_$material.pdf" that was uploaded with the submission})

\subsection{Key Observations}
\label{sec:key_observations}
Salient findings from our analysis are as follows:

\begin{table}
\tbl{Description of Constituent Techniques}{
\centering
\renewcommand{\arraystretch}{1.4}
\begin{tabular}{p{3cm}|p{10cm}}
\hline\hline
\textbf{Constituent Techniques} & \textbf{Specific Variant} \\ \hline

Mathematical Modelling & (a) Logic: Inductive/Deductive Inferences, Logic Frameworks like Fluent Linear Temporal Logic, Boolean satisfiability problem (SAT) solvers, Satisfiability Modulo Theories (SMT) etc. ; (b) Automaton: Context Free Grammar (CFG), Context Sensitive Grammar (CSG), Finite State Machine (FSM); (c) Model Based Design and Verification. \\ \hline

Artificial Intelligence & (a) Machine Learning: Clustering Hypothesis, Learning Algorithms; (b) Search-Based Techniques like hill-climbing, simulated annealing (c) Genetic Programming;  (d) Expert System; (e) Programming By Example; (f) Evolutionary Algorithms.\\ \hline

Human-Computer Interactions & (a) Natural Language Processing tools and techniques. \\ \hline

Probabilistic Modelling & (a) Probabilistic Model; (b) Language Model \\ \hline

Statistical Inferencing & (a) Statistical Model; (b) Statistical Analysis tools  \\ \hline

Programming Paradigms & (a) Symbolic Execution; (b) Object Constrained Language; (c) Other Modeling Languages. \\ \hline

Applied Mathematics & (a) Optimization techniques: Forward and Backward Slicing; (b) Numerical Analysis: Interpolation; (c) Graph Theory: Graph Analysis, Graph Based Modeling, Influence Graph, Decision Tree etc. \\ \hline

Pure Mathematics &  (a) Linear Algebra; (b) Refinement Calculus.  \\ \hline

Correlation-Based Inferencing & (a) Knowledge Base: Ontology; (b) Repositories.  \\ \hline

Information Retrieval & (a) Contextual Search; (b) Clustering Hypothesis  \\ \hline

Domain Engineering & (a) Model Driven Architecture.  \\ \hline
\hline
\end{tabular}}
\label{tab:const_tech}
\end{table}

\begin{table}[]
\tbl{Distribution of articles}{
\centering
\begin{tabular}{|l|l|l|l|l|l|l|}
\cline{1-7}
\textbf{Constituent techniques $\downarrow$} \rotatebox{90}{\textbf{SDLC activity $\downarrow$}}
& \rotatebox{90} {\textbf{Requirement}}
& \rotatebox{90} {\textbf{Architecture}}
& \rotatebox{90} {\textbf{Design}}
& \rotatebox{90} {\textbf{Implementation}}
& \rotatebox{90} {\textbf{Testing$/$Verification}}
& \rotatebox{90} {\textbf{Maintenance}} \\ \hline 

Mathematical Modelling & 11 & 3 & 6 & 16 & 47 & 7 \\ \hline
Artificial Intelligence & 1 & 6 & 1 & 10 & 19 & 8 \\  \hline
Human-Computer Interaction & 2 & 1 & 0 & 3 & 1 & 2 \\  \hline
Probabilistic Modelling & 0 & 0 & 0 & 3 & 2 & 0 \\ \hline
Statistical Inferencing & 1 & 0 & 0 & 0 & 2 & 1 \\ \hline
Programming Paradigms & 0 & 1 & 0 & 1 & 15 & 3 \\ \hline
Applied Mathematics & 5 & 0 & 3 & 2 & 9 & 3 \\ \hline
Pure Mathematics & 0 & 0 & 0 & 4 & 0 & 0\\  \hline
Correlation-Based Inferencing & 5 & 2 & 0 & 7 & 4 & 4 \\ \hline
Information Retrieval & 0 & 0 & 0 & 1 & 4 & 1 \\  \hline
Domain Engineering & 1 & 0 & 0 & 0 & 6 & 5 \\ \hline
\end{tabular}%
}
\label{tab:articles_dist}
\end{table}

\begin{figure}[tbp]
\centering
  \includegraphics[scale=0.7]{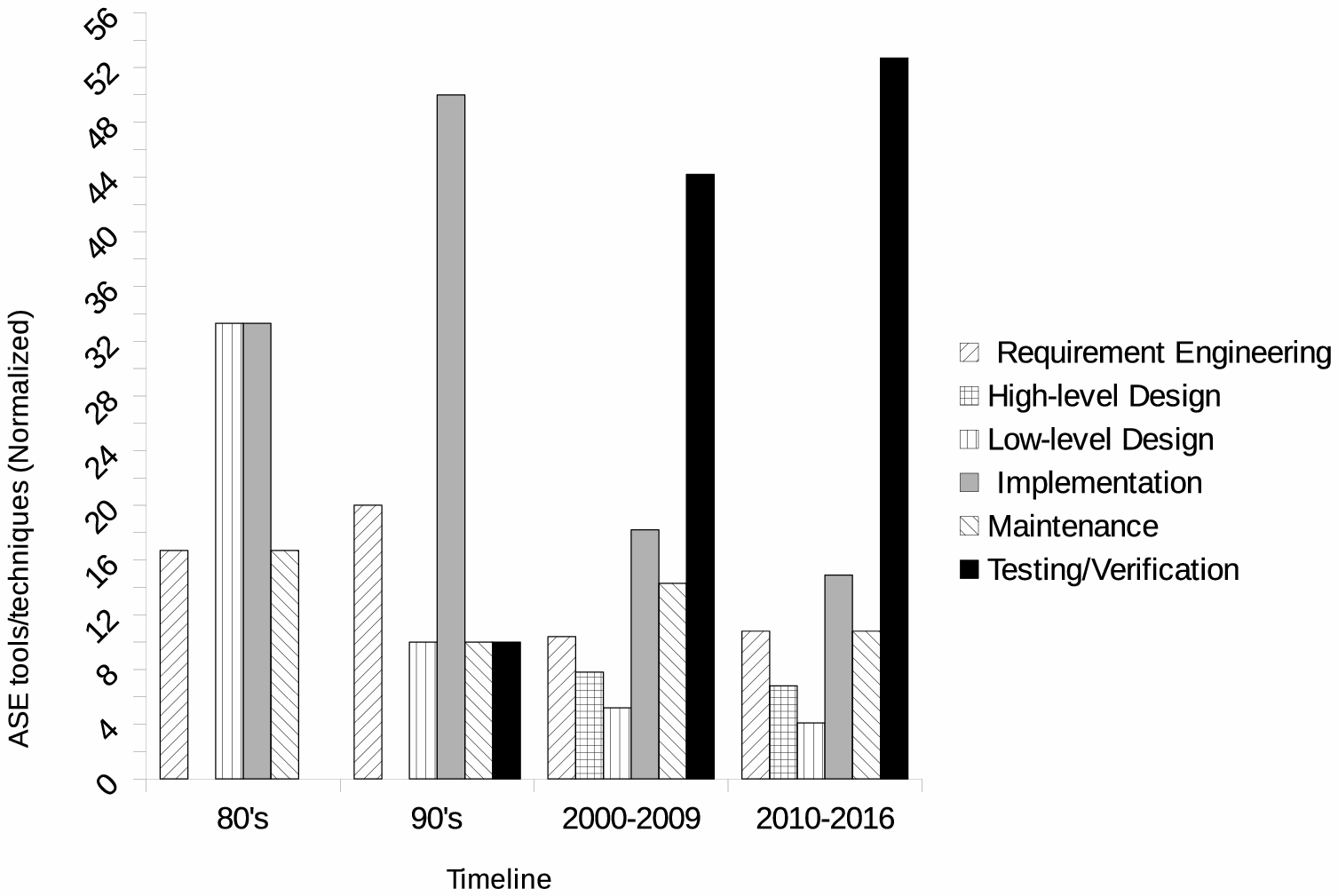}
 \caption{Period vs. Automation in SDLC activities}
  \label{fig:yrwisephase}
\end{figure}

\begin{figure}[tbp]
\centering
  \includegraphics[scale=0.8]{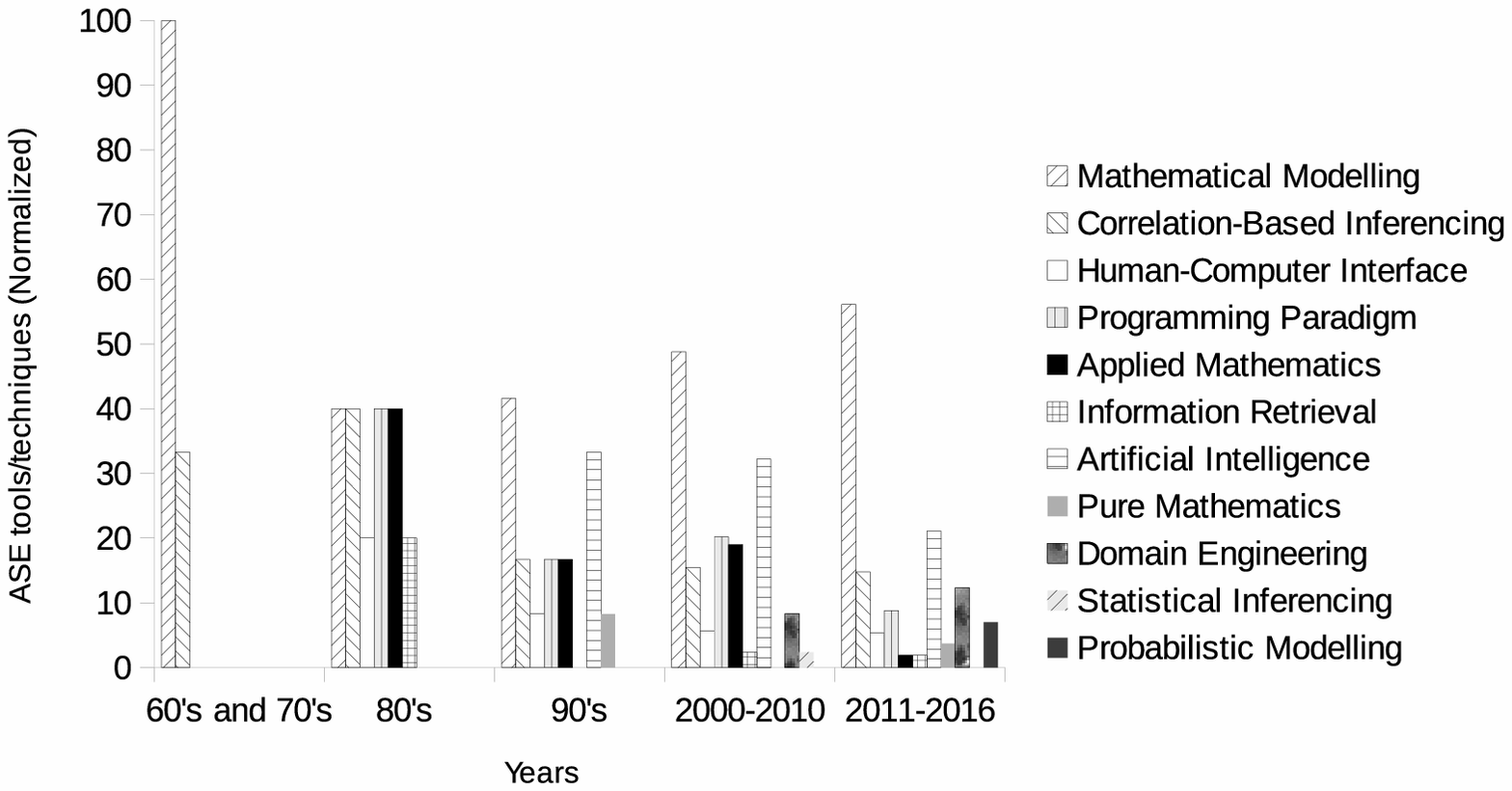}
  \caption{Period vs. Usage of constituent approaches}
   \label{fig:yrwiseconsti}
\end{figure}

\begin{enumerate}
\item Our first major observation is the set of constituent techniques that have been leveraged by the researchers for developing ASE tools targeted for different SDLC activities. We classified the constituent techniques into $11$ categories. These $11$ categories of techniques and the specific  technique under a category that we observed being used in each of $164$ papers are depicted in Table-\ref{tab:const_tech}.

\item Most of the constituent techniques span across all the SDLC activities and have changed over time due to advent of new tools and frameworks and advances in computing technologies. For example, basic logic constructs have been used to hand craft more sophisticated ones for solving ASE issues since early eighties, however, the current trend seem to be to use a pre-existing logic synthesis tools for the purpose. So instead of writing codes for deductive inferences many automation tools use available SAT solvers directly. In addition, newer techniques like evolutionary and genetic algorithms have been used with good results for automating activities in some SDLC phases such as Testing and Maintenance. By success of an automation tool or technique we mean: (i) The concerned ASE tool is able to reduce the manual effort of the corresponding SDLC activity, and (ii) The ASE tool has been further used in academia or industry directly or as an inspiration or basis for better tools.

\item There are some constituent techniques which are suitable/preferred for automating a specific set of SDLC activities. This is due to the type of Input and/or Output expected by different activities. For example, Object Constrained Languages are used to automate architecture and design phase as it is used to define rules for UML. It is not seen to be used in automation of any other activity in SDLC.  

\item A significant number ($\approx74\%$) of ASE tools have only a single constituent technique working as a core approach in achieving the tool's goal. This seems to be in conformity with a common best practise to limit the technology diversity/sprawl in a software solution.

\item Constituent techniques from Mathematical Modelling category are almost universally used by ASE tools across SDLC phases. Artificial Intelligence (AI) is another very popular category. Human-Computer Interaction (HCI), Statistical Inferencing, Probabilistic Modelling, Pure Mathematics and Information Retrieval (IR) are the categories from which fewer techniques have been used in ASE tools.

\item The following SDLC phases, in that order, appear to be favoured the most by ASE researchers: Testing and verification, Implementation, Maintenance, Requirement Engineering. Architecture and Design phases seem to be the least favoured ones. Such popularity may imply the relative ease or difficulty in automating activities in these phases. 

\item Though almost all SDLC phases have seen increased automation over the years, implementation and testing and verification phases have received the most attention during last 10-15 years. Overall, there has been a sudden spurt in ASE contributions from the year 2000 onwards. Fig. \ref{fig:yrwisephase} shows the trend.

\item Over the years Mathematical Modelling remained a popular category to draw from when building ASE tools. AI is another category which has remained popular since 1990s. Techniques from IR category after having seen a dip in 1990s have again started to gain traction among ASE researchers. Other recent categories are Domain Engineering and Probabilistic Modelling. Fig. \ref{fig:yrwiseconsti} shows the trend.

\item We also determined the \emph{impact} of ASE tools that we studied. Fig. \ref{fig:usage_of_techniques} depicts  our findings about impact. A large percentage ($\approx63\%$) of the ASE tools were not used further in any manner. About $17\%$ of them are being used in other software systems or research projects. About $11\%$ of the ASE tools have been viewed as seminal works that lead to further work to develop better ASE tools. For example, [P63]\footnote{Complete list of papers that we studied is available at \url{https://goo.gl/DnJvWy}} proposes Randoop\footnote{https://randoop.github.io/randoop/projectideas.html} which has found extensive usage in other research projects and in some OSS. It uses random testing which is feedback-directed and extended its functionality by using non-deterministic lexical analysis. Similarly, [P28] is a semi-automatic approach to assist in both architecture and design phase. It proposed SPE (Software Performance Engineering) and is considered a pioneering work. [P48] uses model checking for automation and is used in other research projects for code generation.

The \emph{impact} was determined mainly by exploring the citations as well as other types of references (such as on an OSS project which may have used the tool as a basis). We searched Github\footnote{On-line version control repository.}, Apache\footnote{Hosts number of open-source projects} and Google\textregistered \space Search\footnote{On-line search engine} to find their usage in Open Source Software (OSS) and any other systems. 

\begin{figure}[htp]
\RawFloats
\centering
  \centering
  \includegraphics[scale = 0.65]{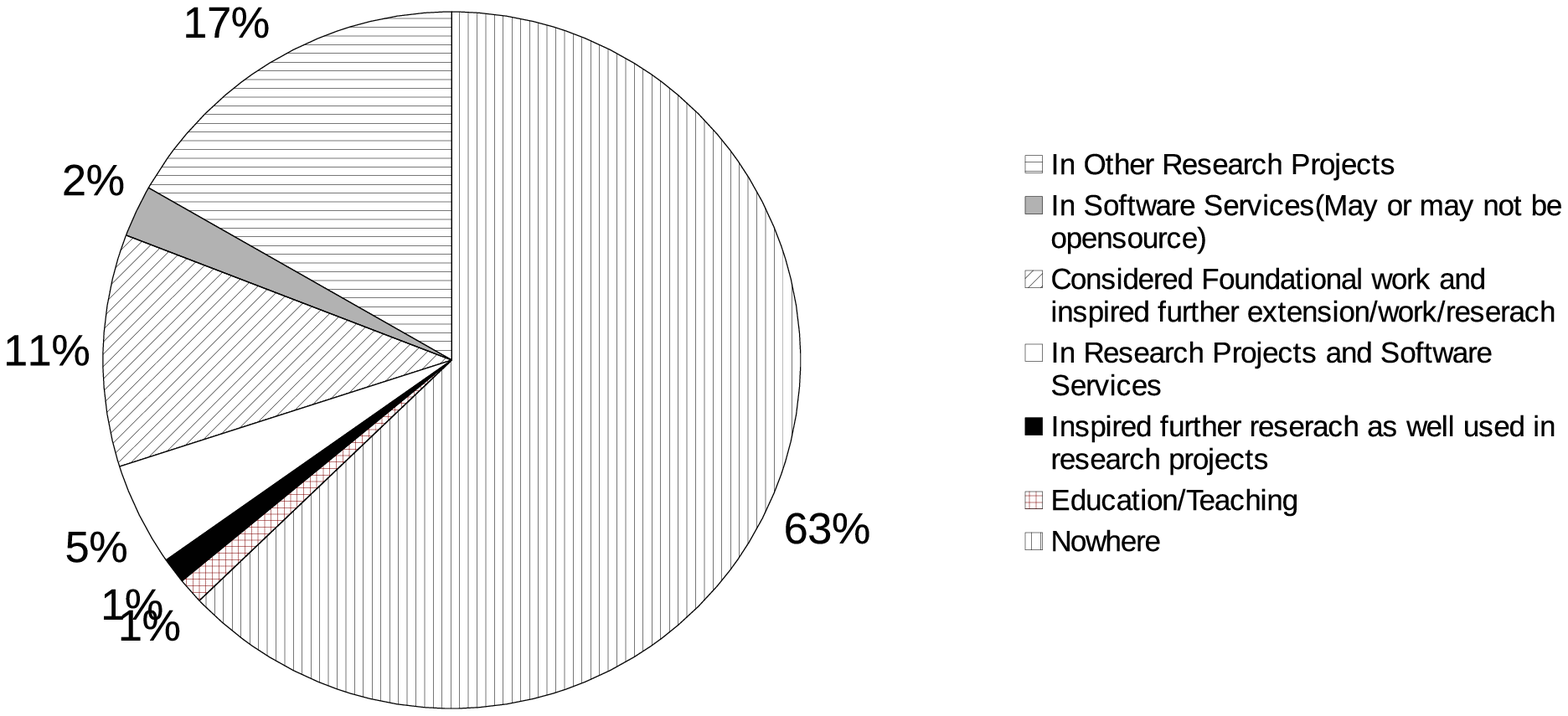}
  \caption{Automation Techniques in Terms of Usage}
  \label{fig:usage_of_techniques}
\end{figure}

\item If we study constituent techniques in isolation from what they are automating then the trends are mostly confusing. This is because a particular constituent technique may have found success in some ASE tool whilst not in others. Form our analysis we inferred that this success or failure is based on what is being automated. For example, evolutionary algorithms when used to automate testing and maintenance activities have shown considerable success than when used for automating requirement analysis. 

\end{enumerate}

\subsection{Observations related to research questions}
\label{sec:res_quest_discussion}
The research questions which we explored are mainly rooted around the following points:
\begin{enumerate}
\item Identify if there exist any relationships between the ASE tools and the issues/activities that these tools addressed.
\item Identify if there exist any relationships between the constituent techniques and the ASE tools that leveraged them.
\item How these relationships evolved/changed with time (i.e. when technology landscape all around advanced).
\end{enumerate}

In preceding section we have highlighted our key observations addressing the above root questions. Here we present the findings which relate to the specific questions that we enumerated in Section-\ref{sec:res_questions_list}.

\textit{RQ.1: } Our findings about patterns of relationships between SDLC activities, the ASE tools and the constituent techniques are depicted in Fig. \ref{fig:yrwisephase}, \ref{fig:yrwiseconsti}, and \ref{fig:usage_of_techniques}. 
	\begin{itemize}
		\item \textit{RQ.1.1: } The constituent techniques that are used to automate different phases of SDLC have been categorised and listed in Table-\ref{tab:const_tech}.
		\item \textit{RQ.1.2: } There is no clear pattern about a constituent technique being ``SDLC activity specific". In other words, it is difficult to say from our findings that a particular constituent technique finds more acceptance in a particular activity of SDLC.
		\item \textit{RQ.1.3: } There are constituent techniques which span across more than one activity of SDLC. Table-\ref{tab:articles_dist} shows the distribution. For example, techniques from Mathematical Modelling category have been used in ASE tools catering to almost all SDLC phases.
		\item \textit{RQ.1.4: } As we have outlined in preceding section, constituent techniques have undergone changes in their reception. Fig. \ref{fig:yrwiseconsti} depicts the observed trend of constituent techniques usage with time.
	\end{itemize}

\begin{figure}
  \centering
  \includegraphics[width=.95\linewidth]{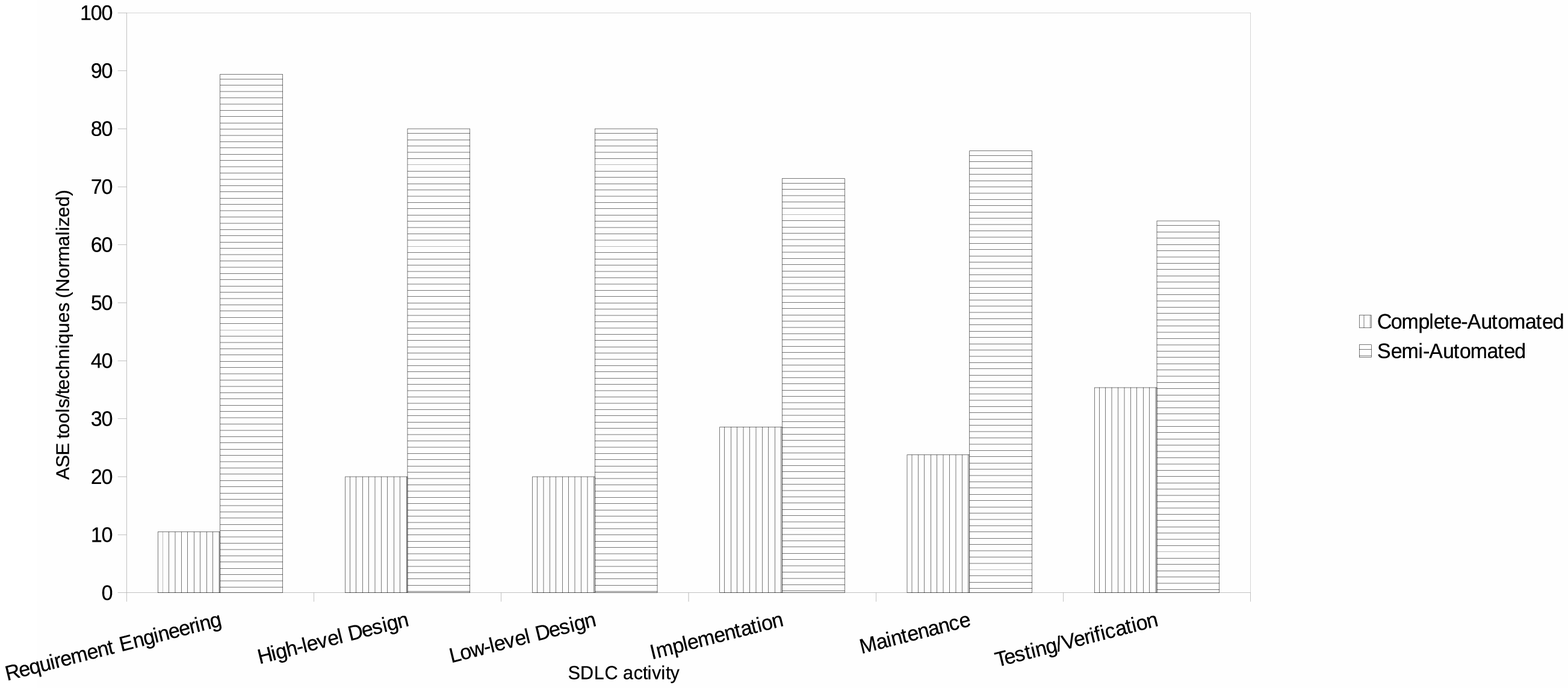}
	\caption{Activity Wise Automation}
  \label{Degreeofa}
\end{figure}

\textit{RQ. 2: To what extent does the automation of a particular SDLC activity take place? Does this extent depend on that activity?}\label{RQ2}
\textit{RQ. 2.1: Is there a particular SDLC activity which is more difficult/easy to automate?}
\\

The difficulty (or ease) of automation of activities in a particular SDLC phase can be deduced from the extent of automation available via ASE tools targeted for such activities. Only about $27\%$ papers in our SLR, proposed an ASE tool which  completely automates the target activities. Other $79\%$ provide semi-automation. Our observations about extents of automation in different SDLC activities are depicted in Fig. \ref{Degreeofa}. The actual number of papers in each activity were converted to percentage of papers for that activity, so as to provide a normalized value. As can be observed, requirement activity has the least number of completely-automated tools and testing the most. 

An ASE tool is said to be offering \emph{complete automation} if it accepts input in the form as is delivered from an earlier SDLC phase and gives output in the form which can be directly fed in to the next phase with minimum intervention from the user. The ASE tools which do not fall in the \emph{complete automation} category are said to provide \emph{semi-automation}.

Whether an ASE tool can be categorized as providing complete or semi automation can be inferred by considering the degree of effort required from a user when using the tool. Such evaluation usually involves examining the format of input and output of the tool. Table-\ref{semiautotabi} and-\ref{semiautotabo} describe the evaluation points used for such categorization.

\begin{table}
\RawFloats
\centering
\setlength\tabcolsep{4pt}
\begin{minipage}{0.45\textwidth}
\centering
\tbl{Classification in Semi-automation Based on Input (In order of increasing user effort)}{
\begin{tabular}{p{1cm} p{4.5cm}} \hline \hline
  \textbf{S.No.} & \textbf{Description of semi-automation} \\ \hline
  1. & User interacts with tool in natural language \\
  2. & Code or Design in some preprocessed form \\
  3. & User has to learn a query language \\
  4. & User has to learn some high-level language  \\ 
  5. & User has to learn some mathematical formalism like First-Order logic\\ \hline
\end{tabular}}
 \label{semiautotabi}
\end{minipage}%
\hfill
\begin{minipage}{0.45\textwidth}
\centering
\tbl{Classification in Semi-automation Based on Output (In order of increasing user effort)}{
\begin{tabular}{p{1cm} p{4.5cm}} \hline \hline
  S.No. & Description of Semi-automation\\ \hline
  1. & Fix the output at right place in that particular phase \\
  2. & User fixes the output at the right place \\
  3. & User has to choose one output (from a list of Outputs) which needs further modification\\
  4. & User intervention (during the running of algorithm) can fine-grain the output/s\\ \hline
\end{tabular}}
\label{semiautotabo} 
\end{minipage}
\end{table}


\paragraph{Is there a particular SDLC activity which is more difficult/easy to automate?} 
If we distinguish in terms of extent of automation we find that most of the early activities of SDLC such as requirement engineering, architecture and design are mostly semi-automated and later activities like implementation, debugging and testing have completely-automated tools{\footnote{With the exception of maintenance. This may be because maintenance requires more manual interventions than all other later activities. To recall, a completely-automated tool will require minimum possible human intervention/effort in addition to Input/Output constraints.}. This makes us aware of a deeper relation between extent of automation and ease of automation of a particular activity. Each automation tool of SDLC activity would have aimed to completely automate it. The lesser it was able to do it suggests the difficulty of automation in that software activity. Thus, as shown in Fig. \ref{Degreeofa}, requirements phase activities seem difficult to automate and those of testing phase appear to be the easiest to automate. We believe this is attributed to the input and output artefacts of these activities. The artefacts related to SDLC phases coming in the beginning of a software development project are found to be more abstract than the artefacts related to later SDLC phases.

\subsection{Threats to Validity}
The domain in which we conducted SLR is huge. The number of proposed ASE tools and techniques in few SDLC phases is enormous\footnote{Surveys exist for SDLC phase specific automation tools.}. To conduct a reliable study in a reasonable time frame, we decided to limit the scope of our search to an authentic and widely accepted on-line repository -- ACM portal -- of research literature that provides latest research in the domain of our interest. This narrowing down may have led to missing of few studies. However, we have tried to make up for it by using reference chaining.

Another limitation could be the exclusion of those ASE tools which were targeted for specific platforms. For example, an ASE tool which assisted in automatic code generation for embedded systems was excluded from our study. This is because we desired to study those ASE tools which were agnostic to specific target platforms. 

\section{SEAT -- A Taxonomy for Software Engineering Automation}\label{tax}
\label{sec:seat}
Appropriate level of abstraction is desirable when describing and organizing scientific knowledge so as to effectively apply such knowledge for addressing practical issues \cite{sjoberg2007future}.
The concepts in a domain and the inter-relationships among them can be structured at several relevant levels of abstraction by arranging such knowledge as a taxonomy. The observations about ASE domain as presented in preceding sections provide a basis for us to develop a taxonomy for automation in software engineering; we call this taxonomy SEAT -- Software Engineering Automation Taxonomy.

\subsection{Motivation and development process}

A taxonomy of automation in software engineering is desirable because, among other uses, it will enable synthesizing new automation tools for different SDLC activities. Re-composing existing systems to achieve better features will also be possible. Such a taxonomy can serve as a semantic tool which will allow for methodical reasoning about various aspects of automation in software engineering tasks. For instance, an ASE researcher can infer the alternatives when identifying constituent techniques to use in an ASE tool that he/she is building. The taxonomy, in addition to being useful for research community, will also serve as an effective aid when teaching automation in software engineering.

A taxonomy can be developed either in a \emph{top-down} approach or in a \emph{bottom-up}. In a top-down approach one requires an existing schema of concepts classification in the concerned domain. The bottom-up approach starts by first capturing the knowledge/identification of existing concepts in the domain for which taxonomy is to be developed \cite{unterkalmsteiner2014taxonomy}. 

We adopted a bottom-up approach when building SEAT. We chose this approach mainly because to the best of our knowledge there wasn't any classification schema which exists for the ASE domain covering all SDLC activities (though few literature surveys in specific SDLC phases exist). As such, a top-down approach would not be suited. The evolution process of SEAT is depicted in Fig. \ref{taxdev}. 

\begin{figure}
  \centering
  \includegraphics[scale=0.4]{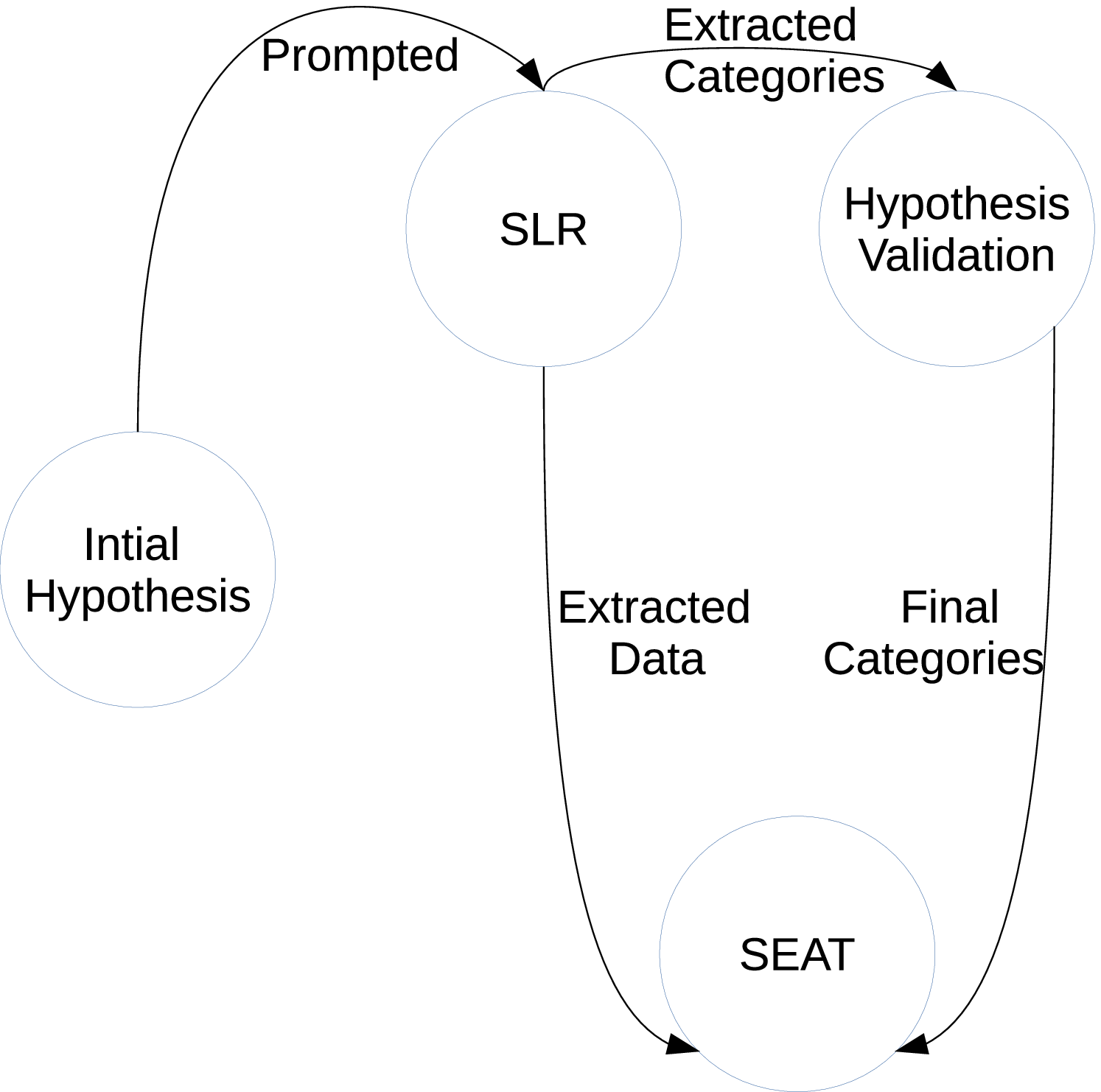}
  \caption{SEAT - The Evolution}
  \label{taxdev}
\end{figure}

We developed an initial hypothesis considering two fundamental aspects of an ASE tool: a) \emph{what} is it that the tool is automating, and b) \emph{how} is such an automation achieved.

The aim was to capture inter-relationships among constituents of ASE approaches, tools and techniques by the \emph{how} dimension. Similarly the \emph{what} dimension was chosen to enable our taxonomy to capture relationships with respect to intended goals of various ASE approaches, tools and techniques. As a next step we performed a systematic literature review (SLR) in the ASE domain. The categories generated by SLR were then used to validate\footnote{Validation was performed on the basis of first research question discussion in Section-\ref{sec:results_discussion}.} our initial hypothesis. The classification with which we began in our hypothesis and the one which we inferred from SLR were found to be largely in sync. As such we used the classification schema which we inferred from SLR to build our taxonomy.

\subsection{Discussion}
SEAT is a two-dimensional, non-orthogonal taxonomy. The overall structure of SEAT is depicted in Fig. \ref{fig:awesome_image}.
\begin{sidewaysfigure}
\includegraphics[scale=0.19,trim=2.3cm 0cm 0cm 0cm, clip=false]{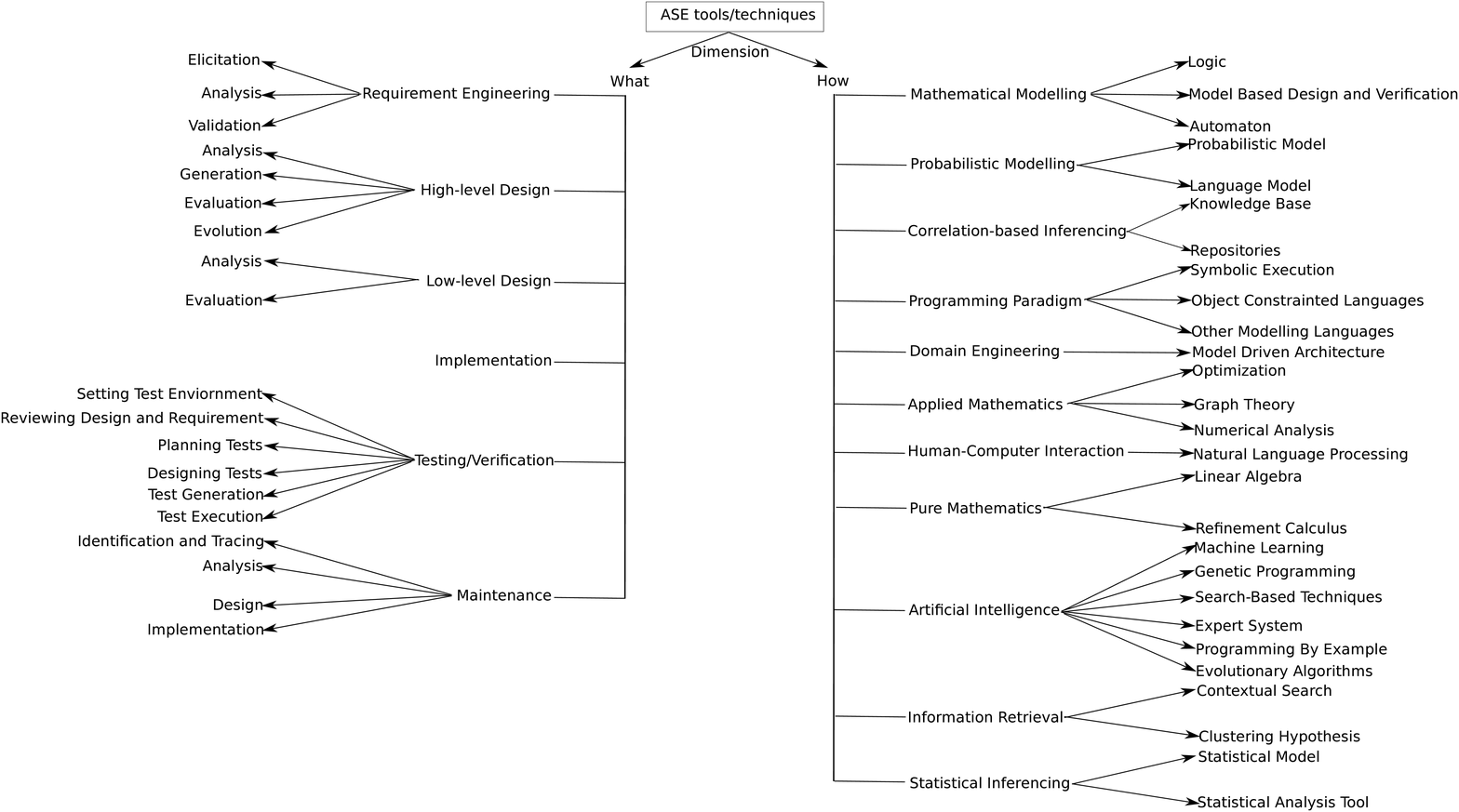}
\caption{Taxonomy of Automation tools for SDLC activities}
\label{fig:awesome_image}
\end{sidewaysfigure}

We observed from our SLR that automation in software engineering has largely been aligned with the fundamental activities of software engineering processes. These activities are: 1) Requirement Engineering 2) Design 3) Implementation 4) Validation (testing/verification) and 5) Evolution of software. Objectives, inputs and outputs of each of these activities are well understood and defined by the software engineering community \cite{bourque1999guide}\cite{Sommerville2010}. Therefore it seems fit to classify various ASE tools along these these activities (i.e. the ``what" dimension).

In Section-\ref{sec:results_discussion} we identified several constituent techniques, from different areas of computing, that have formed the basis of major work in ASE domain. Such constituent techniques have been used to automate different software engineering activities. The identified categories of the constituent techniques are shown in Fig \ref{fig:awesome_image}. 

The abstractions/concepts that we found in both dimensions (i.e. \emph{what} and \emph{how}) are depicted as the second level branches in Fig.-\ref{fig:awesome_image}. We further refine them by identifying the specific SDLC activity which the ASE tool automates (\emph{what} dimension), and also by identifying the specific constituent technique (\emph{how} dimension) that the ASE tool used. Further, the third level of branching in SEAT depicts the set of concept instances that were obtained from our systematic literature review.

An ASE tool may automate more than one activity of SDLC. Also, the ASE tool may make use of more than one constituent technique to achieve automation. For instance, [P162] automates both implementation and testing activity of SDLC using machine learning and a pre-defined language model. In essence, the tool spans two activities of ``what" dimension - namely Software Implementation/Construction and Testing (Level 2, left branch in Fig. \ref{fig:awesome_image}). In addition, two approaches of ``how" dimensions are used - Artificial Intelligence and Natural Language Processing (Level 2, right branch in Fig. \ref{fig:awesome_image}). This makes SEAT a non-orthogonal taxonomy since a particular ASE tool may span across different sub-categories of ``how" and ``why" dimension.
 
\subsection{SEAT In Action}
We observe that SEAT is a directed graph which follows a hierarchy. Therefore we stored it in a graph database for extracting inferences and further synthesis of additional knowledge. We created a graph using neo4j\textregistered \space to depict all nodes and their relationships in SEAT. neo4j is an open-source graph database. It provides all the characteristics of a traditional database, in addition to implementing property-graph model. Each node corresponds to a category of SEAT. The level of node in SEAT is depicted as its property in neo4j to maintain hierarchy. We further populated the graph by creating relationships between leaves that are shown in Fig. \ref{fig:awesome_image}. These inter-leave relationships are based on the actual data obtained from our systematic literature review. The schema of the graph database storing SEAT is depicted in Fig. \ref{schema}. It is basic structure of SEAT and shows various types of nodes in our graph database. 

An advantage of storing the taxonomy in a graph database is that it allows attaching additional data to nodes as well as relationships. For example, in Fig. \ref{schema} we have added two additional properties in the graph: a) \textbf{time\_factor}: A numeric measure that indicates the relative usage of a `constituent technique' in the last $6$ years. b) \textbf{usage\_type}: It is a property that indicates the impact of a constituent technique in terms of influencing/spurring more ASE research.

 \begin{figure}[]
 \caption{Schema (basic structure) of SEAT stored as graph database}
 \label{schema}
 \centering
 \includegraphics[scale=0.13]{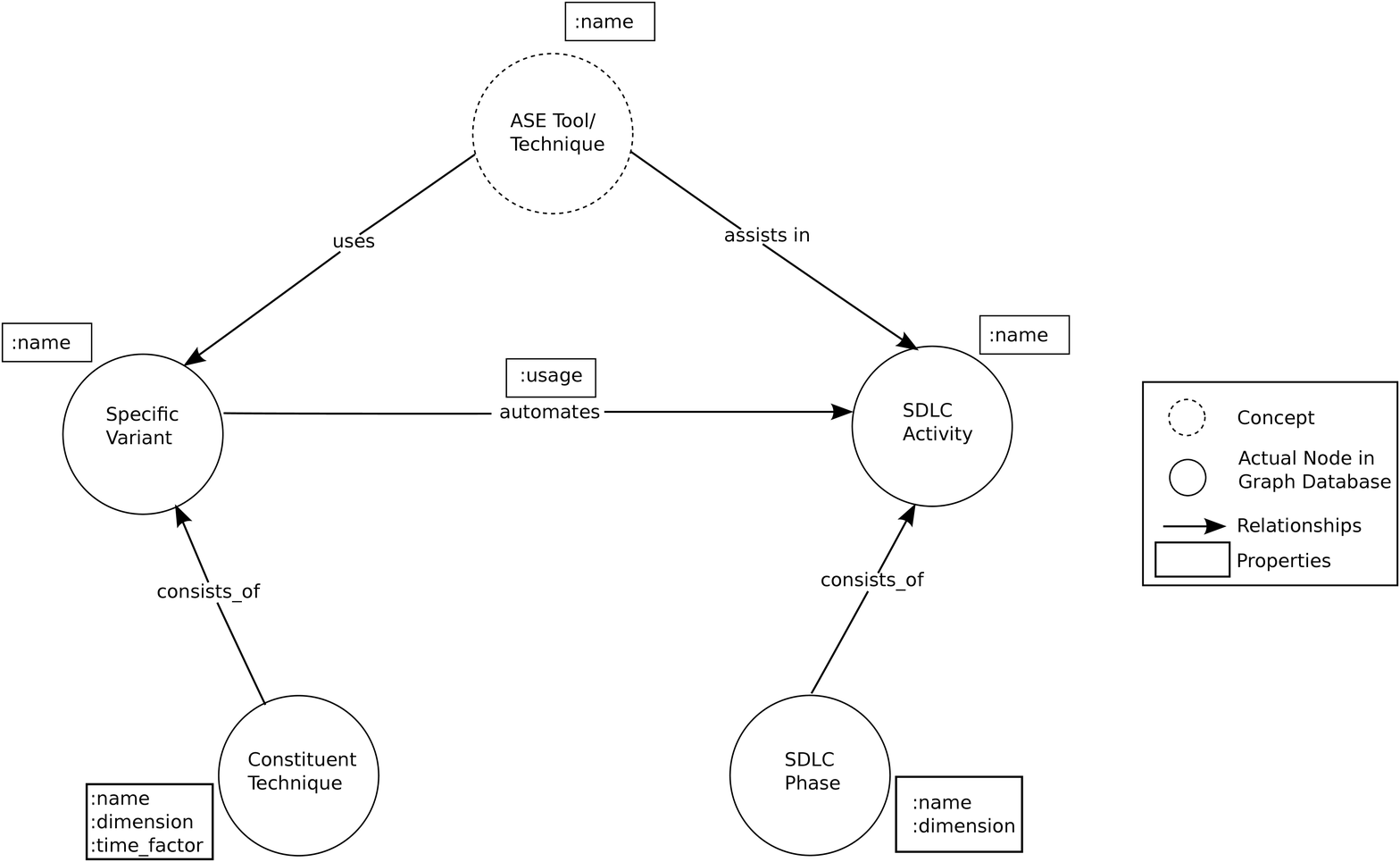}
 \end{figure}

The idea behind storing SEAT in graph is two-folds: a) The graph can further be extended and scaled as more data in this domain is gathered. b) A user can query the graph directly to find the relations between various concepts present in SEAT, instead of manually going through the sizeable information that the taxonomy represents. Also, additional information in the graph helps a user in making useful deductions via graph queries\footnote{neo4j provides a user friendly, SQL like, query language called Cypher (\url{https://neo4j.com/developer/cypher-query-language/#_about_cypher}).}.

In order to test the practicality and utility of SEAT-graph we depict its functioning with three unique and distinct use cases. 
\begin{enumerate}
\item{USE CASE-1:} A basic scenario involves applying SEAT to an ASE tool\footnote{These tools are not part of the SLR we conducted.} for classifying the tool. For instance, a software engineer studies \cite{goldstein2015automatic}-paper and wants to see its placement w.r.t to SEAT. She can run Query-\ref{query1} (written in Cypher) by providing relevant information from the paper in question. Line $1$ and $3$ in Query-\ref{query1} correspond to the process of matching `What' and `How' dimensions with `Architecture Validation' and `Graph' respectively. The $[*1]$ in these statement means that the relationship may traverse more than $1$ path in the graph.
 \begin{lstlisting}[language=SQL,showspaces=false,basicstyle=\ttfamily,basicstyle=\footnotesize,label=query1,caption=Applying SEAT to existing ASE tool,numbers=left,showstringspaces=false]
 MATCH (a)-[*1]->(b) WHERE a.Dimension contains `What' 
 AND b.name contains `Architecture Validation' 
 MATCH (c)-[*1]->(d) WHERE c.Dimension contains `How' 
 AND d.name contains `Graph' 
 RETURN (a)-[]->(b), (c)-[]->(d)
 \end{lstlisting}
 $1^{st}$ row in Table-\ref{query-result} depicts the results of Query-\ref{query1} when executed against our database. ``Software Architecture/High-level Design" is related to `Architecture Validation' in \texttt{What} dimension and ``Applied Mathematics" is related to `Graph' in \texttt{How} dimension.
 \begin{table}[h]
 \tbl{Results of queries against SEAT-graph}{
 \centering
 \begin{tabular}{p{0.2\textwidth} | p{0.7\textwidth}}
 \hline\hline
 \textbf{Query-No.} & \textbf{Results}\\ \hline
 Query1 & (node.name: Software Architecture$/$High-level Design, node.Dimension: What),     (node.Belongs$\_$to: SDLC, node.name: Architecture Evaluation or Architecture Validation) 
 
 (node.name: Applied Mathematics, node.dimension: How, node.time$\_$factor: 3),
 (node.Belongs$\_$to: CT, node.name: Graph Theory) \\ \hline
 Query2 & (node.name: Search-Based Techniques) (rel.name: automates, rel.usage:research$\_$project) (node.name: Design Analysis)
 (node.name: Graph Theory) (rel.name: automates, rel.usage:research$\_$project) (node.name: Requirement Analysis) {\tiny }\\ \hline
 Query3 & (node.name: Mathematical Model, cnt:11), (node.name: Artificial Intelligence, cnt:10), (node.name: Correlation-based Inferencing, cnt:9), (node.name: Domain Engineering, cnt:8), (node.name: Programming Paradigms, cnt:7), (node.name: Probabilistic Modelling, cnt:6), (node.name: Human-Computer Interaction, cnt:5), (node.name: Pure Mathematics, cnt:4), (node.name: Information Retrieval, cnt:3), (node.name: Applied Mathematics, cnt:3), (node.name: Statistical Inferencing, cnt:1)\\ \hline
 \end{tabular}}
 \label{query-result}
 \end{table}
 $1^{st}$ row of Table -\ref{Taxinact} depicts the application of SEAT on \cite{goldstein2015automatic}-paper. Similarly, the graph of SEAT can be used to find application of SEAT to any other existing ASE tool (Table-\ref{Taxinact} is populated with two other similar examples).
\begin{table}[]
\tbl{ASE approaches w.r.t SEAT}{
\centering
\begin{tabular}{p{0.3\textwidth} | p{0.3\textwidth} | p{0.3\textwidth}}
\hline\hline
\textbf{ASE tool} & \textbf{Dimension-What} & \textbf{Dimension-How} \\ \hline
Automatic and Continuous Software Architecture Validation \cite{goldstein2015automatic}& Architecture Evaluation$/$Architecture Validation & Applied Mathematics\\ \hline
Speculative Requirements: Automatic Detection of Uncertainty in Natural Language Requirements \cite{yang2012speculative}& Requirement Analysis & Human-Computer Interaction and Mathematical Modelling\\ \hline
AutoComment: Mining Question and Answer Sites for Automatic Comment Generation \cite{wong2013autocomment} & Maintenance Analysis& Mathematical Modelling and Information Retrieval\\ \hline
\end{tabular}}
\label{Taxinact}
\end{table}

\item{USE CASE-2:} Suppose a software engineering researcher desires to find all constituent techniques that can be used to automate `analysis activity' of an SDLC phase. Further, she is interested in identifying only those constituent techniques which have influenced the development of further ASE tools. Accordingly, Query-\ref{query2} is executed. Line $2$ in Query-\ref{query2} filters the results based on \texttt{usage} property of the relationship.

 \begin{lstlisting}[language=SQL,showspaces=false,basicstyle=\ttfamily,basicstyle=\footnotesize,label=query2,caption= Using SEAT to search for most effective constituent approach,numbers=left,showstringspaces=false]
 MATCH (a)-[r]->(b) WHERE b.name CONTAINS `Analysis' 
 AND r.usage CONTAINS `research' 
 RETURN (a.name),[type(r)],[r.usage],(b.name)
 \end{lstlisting}
 The results, depicted in $2^{nd}$ row of Table-\ref{query-result}, show that ``Graph Theory" when used in automation of \texttt{requirement \textul{analysis}} and ``Search-Based Techniques" in \texttt{\textul{analysis} of software design} has led to ASE tools which have further found usage in other research tools. 

\item{USE CASE-3:} Suppose a researcher is exploring the possibility of developing an ASE tool for assisting in an SDLC activity. She would like to identify the most \emph{recent} constituent techniques which she may consider. Without the SEAT graph, identifying such information would require sifting through a significant volume of literature. However, this task can be quickly done by running Query-\ref{query3} which identifies such a ranked list of constituent techniques. Result of the query is depicted in $3^{rd}$ row of Table-\ref{query-result}. The results show that ``Mathematical Modelling" has been the most extensively used techniques in the last six years and ``Statistical Inferencing" the least.

\begin{lstlisting}[language=SQL,showspaces=false,basicstyle=\ttfamily,basicstyle=\footnotesize,label=query3,caption= Using SEAT acquire ranked list of constituent techniques,numbers=left,showstringspaces=false]
 MATCH (n:Constituent_Technique)                                
 RETURN n.name, (n.time_factor) AS cnt                          
 ORDER BY cnt DESC                                              
 LIMIT 15
\end{lstlisting}

\end{enumerate}

In addition to specific use cases discussed above, we provide examples of few other direct inferences that can be extracted from SEAT-graph. These answers give an idea about the scope of research or tool development that can be done in the domain of ASE.
\begin{itemize}
\item \textbf{Q.} Is there any SDLC activity whose ASE tools although being used in research projects/inspiring further research but never used in any software system (open or otherwise)? 

\textbf{Ans.} Implementation/Code generation
\item \textbf{Q.} What are the constituent techniques which have formed the basis of \texttt{impactful} ASE tools (the tool was further used) in different SDLC phases?

\textbf{Ans.}   Requirement Engineering- Graph Theory \\
     High-level Design- Graph Theory, Logic \\
     Low-level Design- Search-based techniques, Logic \\
     Implementation- Logic, knowledge-base \\
     Testing/verfification- Model Based Design and Verification \\
     Maintenance- Logic and Contextual Search \\
\item \textbf{Q.} In which phase of SDLC an ASE tool is likely to be most \texttt{impactful}?

\textbf{Ans.} Testing/Verification \\

\item \textbf{Q.} Suppose a software enginner desires to find all those constituent techniques which whenever used resulted in an impactful ASE tool.

\textbf{Ans.} Programming Paradigms ($5^{th}$ rank in terms of recent usage), Information Retrieval ($9^{th}$ rank in terms of recent usage).
\end{itemize}
\section{Conclusions}
\label{sec:conclusions}
Enhancing the automation of software development tasks has significant far-reaching implications. Automated software engineering (ASE) is an active area of research that aims to replace much of human programmer efforts by offering automation of such tasks. The work in ASE domain relies on tools and techniques from other diverse areas (e.g. information modelling, semantic computing, NLP etc.) of computing and related disciplines. In order to be effective in ASE domain not only is one required to have a systematic understanding of the ASE tools, but also the constituent techniques that such tools leverage. As such a comprehensive characterisation of ASE tools, their constituent techniques as well as their relationships with the SDLC activities that they automate is highly desirable. One of the major contribution of this paper is to provide such a characterisation.

We have developed a taxonomy called SEAT through an extensive systematic literature review of the ASE domain research articles. About 1175 research articles were collected and after applying methodical filtering and elimination we studied about 167 articles in depth. We found that the predominant constituent techniques which are used in various ASE tools can be grouped into 11 categories (shown in Table-\ref{tab:const_tech}). From our comprehensive study we have been able to identify important relationships among ASE tools and their constituent techniques. We have identified interesting trends and patterns in the way various ASE tools use different constituent techniques.

For example we observe that in case of about 74\% ASE tools there is a single constituent technique which serves as the core idea behind the tool. Similarly, we observed that more than 50\% ASE tools that we sampled targeted activities in the Testing and Verification phase of SDLC. One implication of this observation may be that the SDLC activities where population of ASE tools is relatively low are difficult to automate. We also observed that certain constituent techniques have been consistently relied upon for automating activities across almost all SDLC phases. Mathematical Modelling and Artificial Intelligence are two such example categories. In our study we also examined how the use of various constituent techniques varied over the years, and for what types of SDLC activities. Key observations here are that: a) The automation in SDLC phases has in general seen upward trend with Testing and Verification having more than double the activity than any other phase, and b) Techniques from Information Retrieval category after having seen a dip in 1990s have
again started to gain traction among ASE researchers. We also observed that a significant number ($\approx 63\%$) of ASE tools from our sample were not used further in any manner.

The relationships among ASE tools, their constituent techniques and the SDLC activities, and also the useful properties that we observed formed the basis for the taxonomy (SEAT) that we proposed. SEAT has been realized as a graph database. The graph database allows uncovering hidden relationships via exhaustive search for connections and paths between different nodes (i.e. concepts). We have demonstrated the efficacy of SEAT by discussing few practical use cases. We believe that SEAT will enable better comprehension of the ASE domain and assist in identification of new important research opportunities in the domain of automated software engineering.

\bibliographystyle{ACM-Reference-Format-Journals}
\bibliography{refaga}



\medskip






\end{document}